\documentclass[pra,graphicx,reprint,nobibnotes,superscriptaddress,amsmath,showpacs,aps]{revtex4-1}

\usepackage{amsmath}
\usepackage{amsfonts}
\usepackage{graphicx}
\usepackage{subfigure}
\usepackage{hyperref}
\usepackage{setspace}
\usepackage{lipsum}
\usepackage{float}
\usepackage[euler]{textgreek}
\usepackage[percent]{overpic}
\usepackage{nicefrac}
\begin{document}

\title{Optimal control of photoelectron emission by realistic waveforms}

\author{J. Solanp\"a\"a}
\email[]{janne@spam.solanpaa.fi}
\affiliation{Department of Physics, Tampere University of Technology, Tampere FI-33101, Finland}

\author{M. F. Ciappina}
\email[]{marcelo.ciappina@eli-beams.eu}
\affiliation{Max-Planck-Institut f\"ur Quantenoptik, Garching D-85748, Germany}
\affiliation{Institute of Physics of the ASCR, ELI-Beamlines/HiLASE/PALS, Na Slovance 2, 182 21 Prague, Czech Republic}

\author{E. R\"as\"anen}
\email[]{esa.rasanen@tut.fi}
\affiliation{Department of Physics, Tampere University of Technology, Tampere FI-33101, Finland}

\date{\today}

\begin{abstract}
\noindent Recent experimental techniques in multicolor waveform synthesis allow the temporal shaping of strong femtosecond laser pulses
with applications in the control of quantum mechanical processes in atoms, molecules, and nanostructures. Prediction of
the shapes of the optimal waveforms can be done computationally using quantum optimal control theory (QOCT).
In this work we demonstrate the control of above-threshold photoemission of one-dimensional hydrogen model with pulses feasible for experimental waveform synthesis.
By mixing different spectral channels and thus lowering
the intensity requirements for individual channels, the resulting optimal pulses can
extend the cutoff energies by at least up to 50\% and bring up the electron yield by several orders of magnitude.
Insights into the electron dynamics for optimized photoelectron emission
are obtained with a semiclassical two-step model.
\end{abstract}

\maketitle
{}
\newcommand{\mum}{\textmu m}
\newcommand{\wpscm}{\nicefrac{\mathrm{W}}{\mathrm{cm}^{2}}}

\section{Introduction}

When atoms, molecules, and bulk matter interact with strong and short laser fields
new and peculiar phenomena appear, configuring what nowadays we know as attosecond physics or attosecond science~\cite{krauszreview}. In particular, the so-called above-threshold ionization (ATI) has been a particularly appealing subject in both experimental and theoretical physics. In ATI, an atomic or molecular electron is pulled out to the continuum by the action of the laser electric field and, after a subsequent dynamics, which includes the recollision mechanism, either the electron energy or several components of the electron momentum are experimentally measured (see e.g.~\cite{milosevicreview} for a review about both experimental and theoretical developments). The ATI phenomenon was first observed more than three decades ago by Agostini et al.~\cite{agostini}, and it was established that it occurs when an atom or molecule absorbs more photons than the minimum threshold number required to ionize it, hence the name ATI, leaving the leftover energy being converted to the kinetic energy of the released electron.

With the constant advances in laser technology, it is routine today to generate few-cycle pulses, i.e., laser pulses whose electric field comprises only one or two complete optical cycles, which find an ample range of applications in basic science, for instance, in the control of chemical reactions and molecular motion~\cite{schnurer2000,vdHoff2009}. From a technological viewpoint they are the workhorses in the generation of high order harmonics in atoms and molecules and the creation of isolated extreme ultraviolet (XUV) pulses~\cite{ferrari2010,schultze2007}. In a few-cycle laser pulse the electric field can be characterized
by its duration in time and by the so-called carrier-envelope phase (CEP), defined as the relative phase between the maximum of the pulse envelope and the nearest maximum of the carrier wave. When compared with a multicycle pulse, the electric field of few-cycle pulses changes dramatically its temporal shape with the CEP~\cite{Wittmann2009,kling2008}. From a more fundamental viewpoint, it has been experimentally observed that the CEP plays an instrumental role in high-order-harmonic generation~\cite{nisoli2003}, the emission direction of
electrons from atoms~\cite{paulus2001}, and in the yield of nonsequential double ionization~\cite{liu2004}. Currently, investigations of ATI generated by few-cycle driving laser pulses have attracted so much interest due to the strong sensitivity of the energy and angle-resolved 2D photoelectron spectra to the absolute value of the CEP~\cite{paulus_cleo,sayler}. Consequently, this feature of the laser ionized electron renders the ATI phenomenon as a very valuable tool for few-cycle laser pulse characterization. One of the most widely used techniques to characterize the CEP of a few-cycle laser pulse is to measure the so-called backward-forward asymmetry of the energy-resolved ATI spectrum, from which the absolute value of the CEP can be
directly inferred~\cite{paulus2001}. In addition, nothing but the high-energy region
of the photoelectron spectra appears to be the most sensitive one to the absolute CEP and, consequently, electrons with large kinetic energy are needed in order to characterize it~\cite{paulus2003}.

Recent experiments using plasmon field enhancement have demonstrated that the high-order harmonic generation (HHG) cutoff and ATI photoelectron spectra could be extended further~\cite{kim,sergey}. Plasmonic-enhanced fields appear when a metal nanostructure or nanoparticle is illuminated by a short laser pulse. These fields are spatially inhomogeneous in a nanometric region, due to the strong confinement of the so-called plasmonics 'hot spots' and the distortion of the electric field by the surface plasmons induced in the nanosystem. One should note, however, that a recent controversy about the outcome of the experiments of Ref.~\cite{kim} has arisen~\cite{sivis,Kimreply,sivis2013}. Consequently, alternative systems to the metal bow-tie-shaped nanostructures have appeared~\cite{Kimnew}. From a theoretical viewpoint, however, these experiments have sparked an intense and constant activity~\cite{husakou2011theory,yavuz2012generation,ciappina2012high,yavuz2013gas,yavuz2015h2+,Marcelo12OE,
Marcelo12AA,Marcelo12AAA,Marcelo12JMO,Jose13,Marcelo13A,Marcelo13AR,Marcelo13AP,Marcelo13AA,
Marcelo13LPL,Marcelo14,Marcelo14EPJD,Marcelo15CPC,Marcelo15,Husakou11OE,Husakou14A,tikman2016Ryd, yavuz2016h2pEL}.

An step forward would be to use multicolor waveforms or field transients to drive the ATI phenomenon (for a recent article see e.g. Ref.~\cite{koch2016}). These laser sources present unique characteristics, as noticeable sub-fs changes in the laser electric field~\cite{goulielmakis2008single,lefteris2011}. In addition, a large set of parameters is available to control, with great precision, the shape of the laser electric field. For instance, by manipulating both the amplitude and relative phases of the different \textit{colors}, it would be possible to tailor the laser electric field with an attosecond precision~\cite{hanie2014}.

Clever design of the shape of the laser electric field gives us the ability to control quantum mechanical (QM) processes.
Prediction of the parameters for the waveforms can be achieved either via genetic evolution of the laser parameters
in learning-loop experiments (see, e.g., Ref.~\cite{qoct_review2} and references therein) or via \emph{quantum optimal control theory} (QOCT) simulations~\cite{qoct1,qoct2,qoct_review1,qoct_review2},
where computational tools are used to predict the optimal pulse shapes for a given \emph{target}, i.e., the desired outcome of the QM process.
QOCT has been successfully used to control, e.g., ultrafast strong-field phenomena such as high-harmonic generation~\cite{hhg_solanpaa,hhg_rubio_castro_tddft,hhg_chou},
strong-field ionization~\cite{esa_oct_ionization1,esa_oct_ionization2,esa_oct_ionization3,nikolay_oct_ionization}, and
photoelectron emission~\cite{koch2016}.

However, in ultrafast strong-field physics, there have yet to be any experiments using laser pulses designed with QOCT. This is in contrast to many other fields within the QOCT community (see, e.g., Ref.~\cite{epj2_refref} and the references therein). The reason may be, in part, due to the fact that several previous studies using QOCT in controlling strong-field phenomena produce laser pulses
that are not fully compatible with experimental multicolor waveform synthesis despite several advances in incorporating constraints to QOCT (see, e.g., Refs.~\cite{expr_constraints,qoct_review1} and the references therein).
A recent work by \emph{B. B\'odi et al.} in Ref.~\cite{hhg_new_super} brings QOCT within ultrafast strong-field physics towards predicting experimentally feasible waveforms.
In their work, the total laser electric field is a superposition of four predefined pulses of different colors (channels) that are
obtained from experimental setups. This computational scheme, simulating a single multicolor waveform synthesizer, allows experimental compatibility, in principle.

As in the work by \emph{B. B\'odi et al.}, we present the optimizable pulse as a superposition of component pulses (channels), but do not address any specific light-field synthesizer.
Hence, instead of using channel information from an existing experimental setup, each channel is represented by a single-frequency carrier wave with a Gaussian envelope.
This analytical basis has several advantages: while providing experimentally feasible pulses, we can (1) easily change the channel specifications
and (2) use gradient-based optimization methods if desired. We note that QOCT-schemes representing the field in a basis have been proposed and applied earlier to a variety of systems~\cite{molecular_realistic,fourier_representation,crab,crab_constr,Skinner2010248,periodic_parametrization}.
In contrast to previous methods, in our scheme we aim at compatibility with modern waveform synthesizers for ultrashort strong-field physics. First, we use the most natural analytical basis for pulses produced for such systems,
and second, we add CEP as an optimizable quantity. With respect to physical constraints, our method can enforce arbitrary constraints for the total laser electric field as well as the component channels.

In the next Sections we describe the scheme and use it to optimize multicolor waveforms or field transients for different \textit{targets}, namely the photoelectron yield and/or the ATI energy cutoff. The ultimate goal is to push the limits in the energy conversion, in the sense to reach as energetic electrons as possible, with a given input laser energy.

\section{Optimization scheme}

We describe the total electric field as a superposition of $N$ channels represented as ultrashort pulses consisting of a single-frequency carrier wave with a gaussian envelope.
Each of the channels has their own amplitude $A_i$, carrier frequency $\omega_i$, center time $\tau_i$, carrier-envelope phase (CEP) $\phi_i$, and duration $\sigma_i$, i.e.,
\begin{equation}\label{eqn:pulse_repr}
\begin{split}
\epsilon[\mathbf{u}](t) =\sum\limits_{i=1}^N A_i \cos\Big[\omega_i(t-\tau_i)+\phi_i\Big]&\\
\times \exp\left[-\ln\left(2\right)\left(t-\tau_i\right)^2/\sigma_i^2\right]&,
\end{split}
\end{equation}
where $\mathbf u$ denotes the optimizable parameters. We can choose at will
the number of component pulses $N$, and which parameters are kept fixed and which are optimized.
Note that chirp, and in higher dimensions also polarization, can be easily added to the representation as optimizable parameters.
The total field of Eq.~(\ref{eqn:pulse_repr}) always satisfies $\epsilon(-\infty) = \epsilon(\infty) = 0$
and $|\int \epsilon(t)\,\mathrm{d}t| \approx 0$ whenever $\sigma_{i}\gtrsim 200$ a.u. for wavelengths $<2$ \mum;
If  $\sigma_i\lesssim 200$ a.u.,
one would need to add $\int \epsilon(t)\,\mathrm{d}t = 0$ as a global optimization constraint to have the optimized pulses strictly conform to Maxwell's equations, although we omit this in the following demonstrations.

We test our scheme in a one-dimensional (1D) hydrogen-like atom, but full 3D approaches and multielectronic systems within the single active electron approximation (SAE) could be used. The pulses are optimized to maximize the photoelectron yield and energy.
We take up to $N=3$ channels, each with a fixed frequency
and duration. Thus, the optimizable parameters are the amplitudes $A$, CEPs $\phi$, and time-delays via $\tau$s in line with modern waveform synthesis experiments (see, e.g., Refs.~\cite{hanie2014,lefteris2011,hassan2016}).
The 1D Hamiltonian of our system can be written as
\begin{equation}
\hat{H}=\frac{\hat{p}^2}{2}+ V(\hat{x}) +  \hat{x} \epsilon[\mathbf{u}](t),
\end{equation}
where $V(x) = -1/\sqrt{x^2+1}$ is the soft Coulomb potential. We represent the
system on a real space grid of length $L\approx530$ nm (10,000 a.u.) with a spacing $\Delta x\approx13$ pm (0.25 a.u.).

The time propagation begins from the ground state, and the time evolution is calculated by the exponential
mid-point rule with time step $\Delta t\approx1.2$ as, i.e., \mbox{$\hat{U}(t \to t+\Delta t) \approx \exp\left[-i \Delta t \hat{H}(t+\Delta t/2) \right]$}.
Action of the matrix exponential on the state
is, i.e., $\hat{U} |\psi\rangle$  done using the new algorithm from Ref.~\cite{almohy_higham_2011} as implemented
in \emph{SciPy}~\cite{scipy}.

To target the photoelectron spectrum (PES), we optimize the integral of the PES over some energy range $[E_a,\, E_b]$
(see Ref.~\cite{hhg_solanpaa} for a similar target in HHG).
Calculation of the PES is done according to Ref~\cite{pes}, i.e., we calculate the PES
at the end of the pulse using an energy window technique. This target functional can be written as
\begin{equation}\label{eqn:target}
G[\mathbf u] = \langle \Psi[\mathbf u](T) | \hat{O} | \Psi[\mathbf u](T) \rangle,
\end{equation}%
where
\begin{equation}
\begin{split}
\hat{O} &= \int\limits_{E_a}^{E_b}\,\mathrm{d}E\,\,\text{\texttt{PES}}(E) \\
&=  \int\limits_{E_a}^{E_b}\,\mathrm{d}E\frac{\gamma^4}{\left(\hat{H_0}-E\right)^4+\gamma^4},
\end{split}
\end{equation}%
and $\gamma$ is half of the energy resolution ($\Delta E \approx 0.6$ eV for the cases studied in the present work).

Optimization of this target can be achieved in two ways.
First, we can increase the yield of the photoelectrons (larger integrand values),
or second, as the PES has a sharp cutoff, we can extend the cutoff energy (provided that $E_{b}$ has been set large enough).
In practice, the optimal pulses typically fill both these goals, i.e., we get increase both in the photoelectron yield and in the cutoff energy.

These ingredients are already enough for gradient-free optimization schemes, which we will use in the rest of the paper.
Calculation of the gradient of Eq.~(\ref{eqn:target}) for the pulse representation in Eq.~(\ref{eqn:pulse_repr}) would be trivial following, e.g., Ref.~\cite{qoct_review1}.
However, we found that calculation of an auxiliary wavefunction called the \emph{costate}, at the end of the pulse, was numerically challenging for the chosen target operator. For other methods
of calculating the PES it could be easier to obtain the costate (and hence the gradient).
For instance, calculating the PES as a projection to plane waves would result in the costate at the end of the pulse being just a band-pass filtered final wavefunction.

Optimization of the target is done with \emph{Multi-Level Single-Linkage} (MLSL) global optimizer~\cite{mlsl}.
The MLSL algorithm conducts a series of local optimization searches within a bounded domain while avoiding (1) repeated searches of previously found local maxima
and (2) starting local searches near the search space boundaries~\cite{mlsl,nlopt}. For local optimization, we employ a gradient free algorithm called \emph{Bound Optimization by Quadratic Approximation} (BOBYQA~\cite{bobyqazhang,bobyqasource})~\cite{bobyqa} which, together with MLSL, allows us to bound the optimization variables.
In particular, the amplitude of each channel is capped to $A\in[0.03, 0.13]$ a.u., the full-width half maximum (FWHM) to $\sigma\in[3.6, 9.7]$ fs and the maximum time-delay between channels to $\sim$ $9.7$ fs.
The pulse constraints (fluence and peak intensity) are nonlinear in the search space and can not be handled by bounds for the optimization variables. These global constraints are enforced via the augmented lagrangian technique~\cite{auglag1,auglag2}. For the optimization algorithms, we use the \emph{nlopt} library~\cite{nlopt} implementations.

The optimization routine begins from a random pulse configuration
usually giving low yield and small cutoff energies for the PES.
During the optimization, the algorithms find several locally optimal
pulses for our target. Here we show the best of the locally optimal pulses and compare it with:
\begin{itemize}
    \item[i)] a commonly available reference pulse with carrier wavelength of 800 nm with the same fluence and peak intensity as the optimized pulse and
    \item[ii)] the separate channels of the optimized pulse.
\end{itemize}%

\section{Optimization results}

The easiest way to increase the photoelectron energies would be to increase the peak intensity or the wavelength of the driving
laser pulse. There is, however, a limit to the dominant wavelength of strong femtosecond laser pulses, and currently, in experimental multicolor waveform synthesis, it is easier to distribute energy between different channels than to concentrate it all
to a single channel~\cite{hanie2014}.

\begin{figure}[t]
\includegraphics[width=\linewidth]{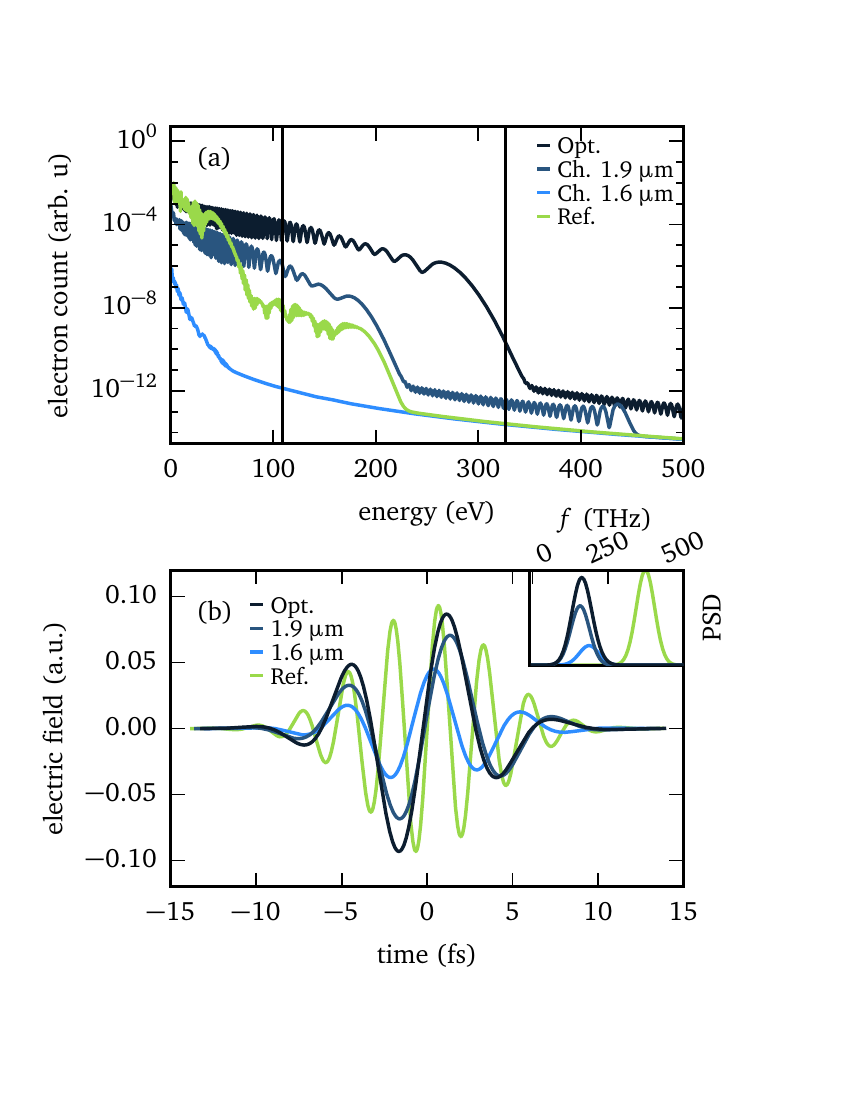}
\caption{(a) Optimized photoelectron spectrum (PES) with a two-channel pulse (black) demonstrates
up to 3 orders of magnitude increase in the yield and over 100 eV extension of the cutoff energy compared to the nm single-channel ($800$) reference pulse [green (light gray) curve], and single channels of the optimal pulse [blue lines].
(b) The optimized pulse (black) is composed of two channels (dark blue and blue),
and the reference pulse (green) has the same peak intensity and fluence as the optimized pulse, but different spectral range.
The power spectral distributions of the pulses are shown in the inset.}
\label{fig:N2}
%The intensities of the single channels for the optimized pulse are $1.9\cdot10^{13}~\wpscm$ and $5.9\cdot10^{14}~\wpscm$.
\end{figure}

Hence, we begin by setting up two spectral channels, the simplest possible multicolor waveform configuration.
The channels have partially overlapping spectral shapes with central frequencies corresponding to wavelengths of 1.6 \mum\ and 1.9 \mum.
Furthermore, the peak laser electric field is constrained below $0.09$ a.u. (corresponding to a peak intensity $\approx 2.8\cdot10^{14}$ W/cm$^2$),
and the fluence to $3$ a.u., but it turns out that the peak field constraint is more restricting than the fluence constraint in this case.
The targeted energy range is approximately from $E_{a}=110$ eV ($\sim 4$ a.u.) to $E_{b}=330$ eV ($\sim 12$ a.u.) shown as vertical lines together with the spectra in Fig.~\ref{fig:N2}(a).

The optimized spectrum [black curve in Fig.~\ref{fig:N2}(a)] has a cutoff energy of $\sim$ 300 eV. This is $\sim$ 50 \% more than for
the $1.9$ \mum\ channel of the optimized pulse (dark blue curve), and in addition, the yield is increased by up to 3 orders of magnitude.
If we compare the optimized spectrum to what is obtained for a commonly available 800 nm pulse (green curve), we observe
even more dramatic enhancements.

The optimized pulse [black line in Fig.~\ref{fig:N2}(b)] mixes the 1.6 \mum\ and 1.9 \mum\ channels roughly in proportions of one to three when comparing their respective intensities.
Essentially, the optimization algorithm finds the correct CEP and time-delay for each channel in order
to increase the peak intensity and fluence of the total field
compared to the 1.9 \mum\ channel only.
This achieves the desired effect, i.e.,
the enhancement of the PES without concentrating all the pulse energy to a single channel.

\begin{figure}
\includegraphics[width=\linewidth]{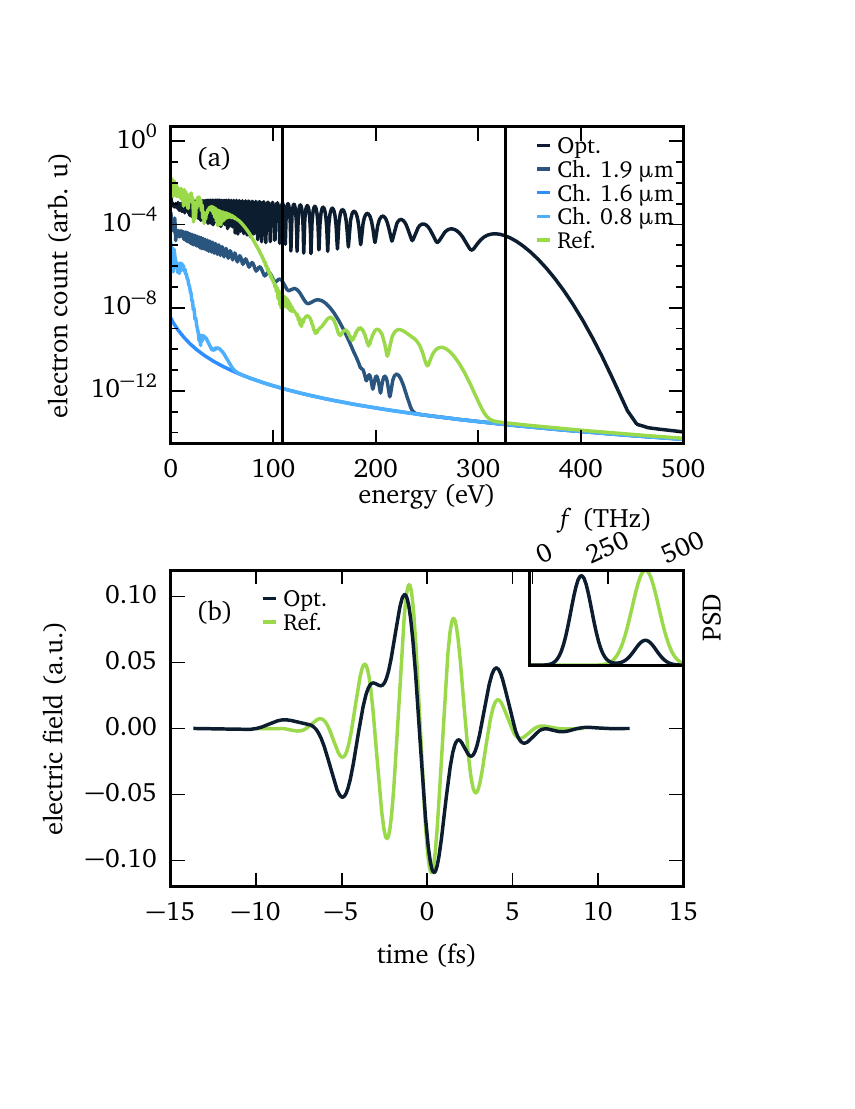}
\caption{(a) Optimized photoelectron spectrum (PES) with a three-channel pulse (black) shows
a yield enhancement up to 6 orders of magnitude and a dramatic cutoff extension (more than 100 eV)
compared to the single-channel ($800$ nm) reference pulse (green), and single channels of the optimal pulse (blue lines).
(b) The optimization changes the dominant spectral contribution to lower
frequencies as seen from the power spectral densities (PSDs) of the laser pulses in the inset,
and it also increases the duration of the major cycle of the optimized pulse.}
\label{fig:N3}
\end{figure}

The pulse shapes allowed by the two-channels are quite restricted, and we can increase the degrees of freedom by adding in another spectral channel. The three-channel optimization is conducted with central frequencies of the channels corresponding to wavelengths 0.8 \mum, 1.6 \mum, and 1.9 \mum,
and we also increase the peak field constraint to $0.11$ a.u. The target remains the same as for two-channel optimization, i.e., from $E_a=110$ eV to $E_b=330$ eV.
Figure~\ref{fig:N3}(a) shows the optimized PES (black) and compares it to the reference spectrum of the 800 nm pulse (green)
and the spectra obtained for single channels of the optimized pulse (blue lines). The optimal pulse increases the
yield up to six orders of magnitude and extends the cutoff energy by over 100 eV,
and as in the two-channel case, the optimal pulse wins over the single channel results.

The optimal three-channel pulse [black line in Fig.~\ref{fig:N3}(b)] mixes
the 0.8 \mum, 1.6 \mum, and 1.9 \mum\ channels in (intensity) proportions of around 5-1-11.
This lowers the intensity requirement for the long-wavelength channels as in the two-channel setup described above.
It is of interest to note that the changes in the PES are due
to mixing lower-wavelength channels, channels which alone give spectra with much lower energy cutoffs.

\begin{figure}[t]
\includegraphics[width=\linewidth]{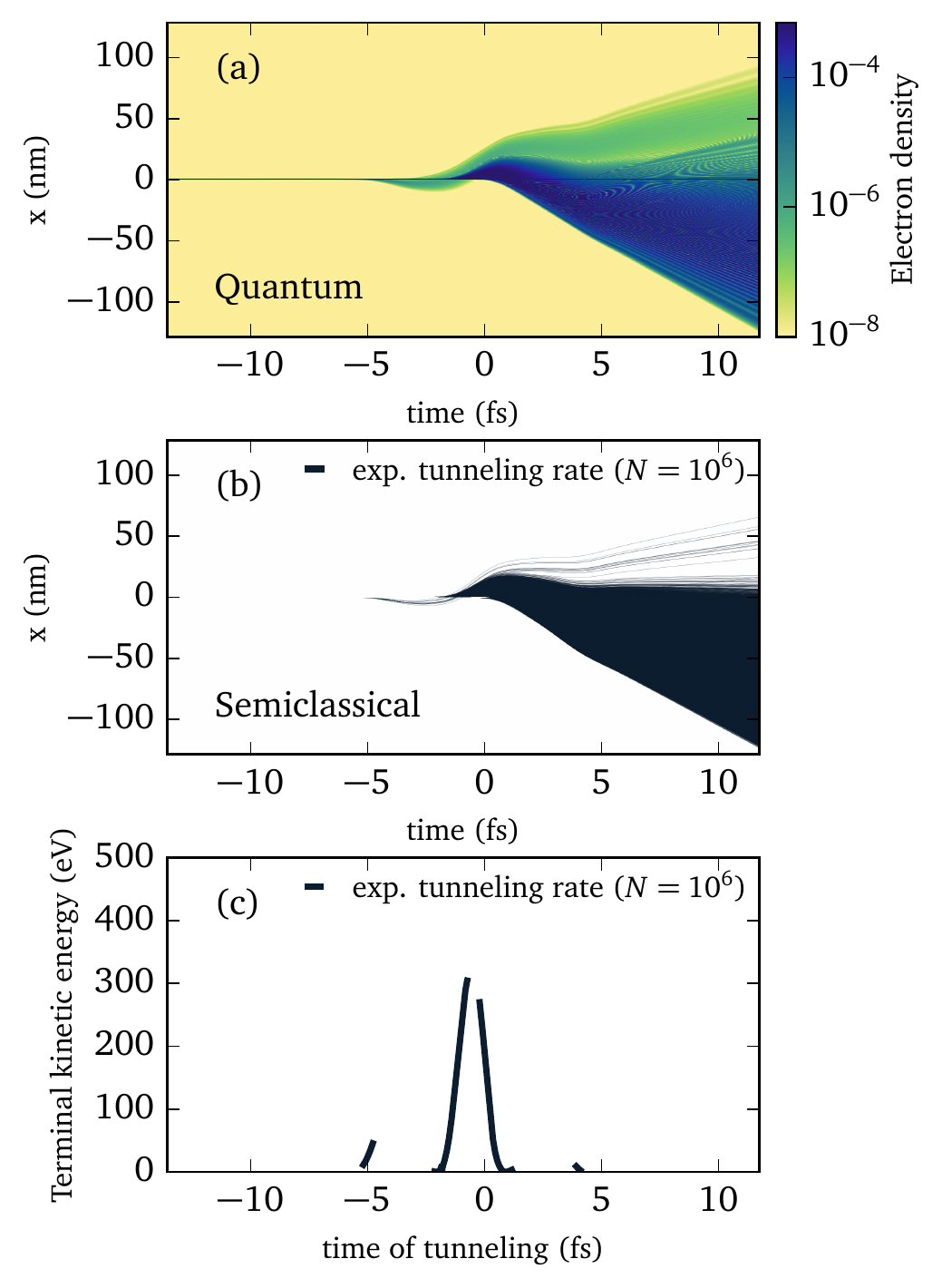}
\caption{(a) Electron density corresponding to the optimized 3-channel pulse in Fig. 3,
(b) $10^{6}$ corresponding (semi)classical trajectories, and
(c) the terminal kinetic energy as a function of the time of tunneling.}
\label{fig:semiclassical}
\end{figure}

One of the advantages of the 1D models is the possibility to scrutinize the time and spatial electron dynamics in a direct way.
To this end in Fig.~\ref{fig:semiclassical}(a) we show the electron density $|\Psi(x,t)|^2$, where $\Psi(x,t)$ is the spatio-temporal electron wavefunction,
for the optimal three-channel laser pulse used in Fig.~3(b). Essentially, the last dominant cycle in the laser pulse
packs as much energy in the ejected electron wavepacket as possible. To better illustrate this idea, we employ a semiclassical two-step model similar to the three-step model used in Ref.~\cite{hhg_solanpaa}.

An ensemble $10^6$ trajectories is simulated as follows:
\begin{enumerate}
\item The tunneling times (start times of the trajectories) $t_0$ are randomized following
the exponential tunneling rate~\cite{PPT, *[{English: }]PPTen, ADK, *[{English: }]ADKen, KrainovJOSAB}
\begin{equation}
w\left(t_0\right)\sim\exp\left\{-\Big[2\left(2I_p\right)^{3/2}\Big]/\Big[3\lvert\epsilon(t_0)\rvert\Big]\right\},
\end{equation}
where \mbox{$I_{p}=0.669$} a.u. is the ionization potential of our system.
\item The trajectories start with zero velocity at the tunnel exit, which is located at
the classical turning point on the farther side of the tunneling barrier.
\item After tunneling, the trajectories are propagated classically, i.e., following Newton's equations, using the 8th order Dormand \& Prince algorithm with adaptive step size control (see e.g.~\cite{rk4}).
\end{enumerate}

Figure~\ref{fig:semiclassical}(b) shows tracing of the classical trajectories. We see a clear correspondence to the QM simulation in Fig.~\ref{fig:semiclassical}(a): quantum mechanically high-density areas are filled with semiclassical trajectories whereas QM low-density areas have only few trajectories. By increasing the ensemble size by a few orders of magnitude, better agreement with the QM low-density region could be obtained.

In the semiclassical model, the maximum kinetic energy is obtained for tunneling events between the two dominant subcycles of the pulse i.e., slightly before $t_{0}=0$ as shown in Fig.~\ref{fig:semiclassical}(c) illustrating the terminal kinetic energy as a function of the tunneling time of the trajectory.
For the trajectories tunneling out near $t_{0}=0$, the semiclassical model yields the maximal terminal kinetic energy of $315$ eV, i.e.,
at the beginning of the cutoff of the optimized QM spectrum of Fig.~\ref{fig:N3}(a).
The electron trajectories that tunnel out near the field minimum at $t_{0} \approx 0$ feel only the full effect of the later dominant half cycle of the pulse, thus contributing to the cutoff region of the PES.
An electron tunneling out earlier would be slowed down by the previous half-cycle, and an electron tunneling out later would not obtain
the maximum energy from the latter half-cycle.

\section{Summary}
We have presented a computational optimal control scheme that composes experimentally feasible multicolor waveforms from analytical pulse components (channels).
As a case study we apply the scheme to
the optimization of photoelectron spectra in a one-dimensional hydrogen-like system. The scheme provides substantial yield enhancement and cutoff extension
compared to single 800 nm pulses with the same peak intensity and fluence or the component channels of the optimized pulse.
By mixing a few different spectral channels, the proposed method decreases the need for high intensities in single spectral channels. Simultaneously, the scheme provides significant
enhancements in the photoelectron spectrum yield and cutoff. In addition, we have shown that the physical working mechanisms behind the optimal pulses can be inspected with simple semiclassical models.

With the chosen channel configurations and target energies
the scheme already provides photoelectrons with $\sim 0.5$ keV energies.
By suitable modifications in the channel
configuration and pulse constraints, the scheme could
provide a way to generate ultrashort electron pulses with
sufficient yield even in the keV regime.
Such electron pulses can be used for diffraction experiments, and could provide an alternative
method to the celebrated laser-induced electron diffraction (LIED) technique (see, e.g., Refs.~\cite{lied_meckel,pullen1,pullen2}), but with a much
finer spatial resolution.

In addition, an extension of our optimal control scheme with realistic waveforms to 3D and many-electron systems is straightforward.
The scheme can be modified to use most of the existing optimization algorithms to account for different search space landscapes in other systems,
and it can be readily be implemented in existing optimal control software or
as an external module to all state-of-the-art software packages for single- or many-electron simulations.
This provides straightforward access to a multitude of different applications including, e.g., optimization of high-harmonic generation, atomic transitions between states, and electron dynamics in molecular and nanoscale devices.

\begin{acknowledgments}
This work was supported by the project ELI--Extreme Light Infrastructure--phase 2 (CZ.02.1.01/0.0/0.0/15\_008/0000162 ) from European Regional Development Fund and by the Academy of Finland (project no. 126205) and COST Action CM1204 (XLIC). We also acknowledge CSC the Finnish IT Center for Science for computational resources.
Several Python-extensions~\cite{[{iPython: }]ipython,[{matplotlib: }]matplotlib,
[{Scipy and Numpy: }]scipy,*scipy2,cmocean} were used for the simulation and analysis.
\end{acknowledgments}

\bibliography{refs}

\end{document}